\documentclass[runningheads]{llncs}
\usepackage{graphicx,comment}
\usepackage{booktabs} 

\usepackage[utf8]{inputenc} 
\usepackage{amsfonts,amssymb,amsmath}
\usepackage{url}
\newcommand{\toto}{xxx}

\newenvironment{proofL}{\noindent{\bf
		Proof }} {\hspace*{\fill}$\Box_{Lemma~\ref{\toto}}$\par\vspace{3mm}}

\newcommand{\vir}[1]{``#1''} 

\usepackage[normalem]{ulem}
\newcommand\redout{\bgroup\markoverwith
	{\textcolor{red}{\rule[0.5ex]{2pt}{0.8pt}}}\ULon}

\newcommand \nil{\textit{nil}}
\newcommand \Fig{\ensuremath{\text{Algorithm}}}

\usepackage{algorithm}
\usepackage{rounddiag}

\usepackage{technote}
\usepackage{homodel}

\newconstruct{\FOREACH}{\textbf{for each}}{\textbf{do}}{\ENDFOREACH}{}

\newcommand\From{\textbf{from}}
\newcommand\Broadcast{\textbf{broadcast}}

\newident{noop}
\newconstruct{\UPON}{\textbf{upon}}{\textbf{do}}{\ENDUPON}{}

\newcommand\Proposal{\mathsf{PRE-PROPOSE}}

 \newcommand\Prevote{\mathsf{PROPOSE}}
\newcommand\Precommit{\mathsf{VOTE}}

\newcommand\hb{\mathsf{HeartBeat}}

\newcommand\coord{\mathsf{proposer}}
\newcommand\sendBy{\mathsf{sendByProposer}}

\newcommand\currEpoch{e}

\newcommand\timeoutPropose{\texttt{timeoutPrePropose}}
\newcommand\timeoutPrevote{\texttt{timeoutPropose}}
\newcommand\timeoutPrecommit{\texttt{timeoutVote}}

\newcommand\timerPrePropose{\textit{timerPrePropose}}
\newcommand\timerPropose{\textit{timerPropose}}
\newcommand\timerVote{\textit{timerVote}}

\newconstruct{\FUNCTION}{\textbf{Function}}{\textbf{:}}{\ENDFUNCTION}{}

\newcommand{\tr}[1]{}
\renewcommand{\tr}[1]{#1}

\begin{document}
\title{Dissecting Tendermint}
%
%
\author{Yackolley Amoussou-Guenou\inst{1,2} \and
Antonella Del Pozzo\inst{1} \and
Maria Potop-Butucaru\inst{2} \and
Sara Tucci-Piergiovanni\inst{1}}
\authorrunning{Y. Amoussou-Guenou et al.}
%
\institute{CEA LIST, PC 174, Gif-sur-Yvette, 91191, France \and
	Sorbonne Université, CNRS, LIP6, F-75005 Paris, France
}
\maketitle              
\begin{abstract}
In this paper we analyze Tendermint, proposed in \cite{BKM18}, one of the most popular blockchains based on PBFT Consensus. Our methodology consists in identifying the algorithmic principles of Tendermint necessary for a specific system model. The current paper dissects Tendermint under two communication  models: synchronous and eventually synchronous ones.  
 This methodology allowed to identify bugs  in preliminary versions of the protocol   and to prove its correctness under the most adversarial conditions: an eventually synchronous communication model under Byzantine faults. 
 The message complexity of Tendermint is $O(n^3)$. 

\end{abstract}
\section{Introduction}\label{sec:intro}
A blockchain is a distributed ledger implementing an append-only list of blocks chained to each other, it serves as an immutable and non repudiable ledger in a system composed of untrusted processes. The append operation needs to preserve the chain shape of the data structure, leading to the necessity to have a mechanism allowing  processes to agree on the next block to append. Bitcoin blockchain, for example, employs the proof-of-work mechanism \cite{DworkN92}, that  is, processes willing to append a new block have to solve a crypto-puzzle and the winning process will append the new block. While this mechanism does not require a real coordination between the processes participating to the Bitcoin system, it might lead to inconsistencies. Indeed, if more than one process solves the crypto-puzzle to extend the same last block then processes may have blockchains with different suffix as long as the conflict is unsolved.

 In blockchain systems area the recent tendency is to privilege solutions based on distributed agreement than proof-of-work. This is motivated by the fact that the majority of proof-of-work based solutions such as Bitcoin or Ethereum are energetically not viable when efficiency is targeted. Moreover proof of work solutions guarantee the existence of an unique chain only with high probability which is the major drawback for using blockchains in industrial applications. That is, forks even though they are rare do still happen with an impact on the consistency guarantees offered by the system and consensus algorithms play an important role to prevent inconsistencies.  In \cite{btadt} the authors proved that consensus \cite{LSP82} is necessary in order to avoid forks.
Therefore, alternatives to proof-of-work have been recently considered and interestingly, the research in blockchain systems revived a branch of distributed systems research: Byzantine fault-tolerant protocols having PBFT consensus protocol as ambassador. 
It should be noted that PBFT solutions cannot be used in permissionless settings if the number of participants to the agreement is not known in advance. That is, in permissionless settings, for each block, a subset of processes (called validators in Tendermint) runs a Byzantine fault-tolerant consensus algorithm to propose the next block to be appended to the blockchain. All the existing solutions for PBFT consensus use the number of validators as hardcore information in their algorithm.

\paragraph{Related Work.}
In the blockchain realm, there exist several Byzantine Fault Tolerant Consensus based blockchain proposals (e.g.,  \cite{solidus,hyperledger,dbft,PeerCensus}, and \cite{BizCoin}).  

The consensus problem, as proved in the seminal FLP paper \cite{flp},   cannot be solved in an asynchronous message-passing system (when there are no upper bounds on the message delivery delay) in the presence of one faulty (crash) process.
Moreover, in \cite{LSP82}, the authors prove that consensus cannot be solved in presence of $f$ Byzantine faulty processes if the overall number of processes $n$ is less than $3f+1$ in a synchronous message-passing system (where the message delivery delay is upper bounded).   In between those impossibility results, it is still possible to solve consensus in an asynchronous setting, either adding randomness \cite{ben83} (which also proved the impossibility result for $n\leq 3f$ for any asynchronous solution) or partial synchrony as in Dwork et al. \cite{DLS88} (DLS) where BFT Consensus is solved  an eventual synchronous message-passing system (there is a time $\tau$ after which there is an upper bound on the message delivery delay). DLS preserves safety during the asynchronous period and the termination only after $\tau$, when the message transfer delay becomes bounded. The  message complexity of this protocol is $O(n^4)$ per epoch and it needs $O(n)$ epochs before deciding. 
Finally, Castro and Liskov proposed PBFT \cite{PBFT99}, a leader-based protocol that optimizes the performances of the previous solution.  If the leader is correct the complexity boils down to $O(n^2)$. Otherwise, a view change mechanism takes place, to change the leader and resume the computation.
The view-change is used to avoid that, in case of faulty leader, if some correct process decides on a value $v$, the other correct processes cannot decide on a value $v' \neq v$ when the new leader  proposes a new value.
Such mechanism implies that when a leader is suspected to be faulty, all processes have to collect enough evidences for the view-change. That is, the view-change message contains at least $2f+1$ signed messages and these messages are sent from at least $2f+1$ processes which yields a message complexity of $O(n^2)$. These messages are then sent to all processes, the view-change has then $O(n^3)$ message complexity. Since the protocol terminates when there is a correct leader, which may happen for the first time in epoch $f+1$, then in the worst case scenario it has a message complexity of $O(n^4)$.
Interestingly,  Tendermint as well as  similar recent approaches e.g \cite{hotstuff} use an alternative mechanism for leader replacement that allows to drop message complexity to   $O(n^3)$. Basically, processes instead of exchanging all the messages they already delivered (used previously to trigger a view change), locally keep track of potentially decided values.
\paragraph{Our Contribution.} In this paper we analyze Tendermint proposed in \cite{BKM18} as one of the most promising but not fully analyzed blockchain protocols   that implements Byzantine fault tolerant consensus. Tendermint targets an eventual synchronous system \cite{DLS88}, which means that safety has to be guaranteed in the asynchronous periods and liveness in synchronous ones, when a subset of processes can be affected by Byzantine failures. 
To analyze the protocol, we dissect Tendermint identifying the techniques used to address different challenges   in the considered system model: synchronous round-based communication model  and eventual synchronous communication model. For each type of model we provide the corresponding algorithm (a variant of Tendermint \cite{BKM18}) and compute its complexity. 
Interestingly,  and contrary to the classical view-changed based approaches,  message complexity in the worst case scenario is $O(n^3)$. This is because processes, instead of exchanging all the messages they already delivered, locally keep track of potentially decided values to preserve the safety, hence reducing the message complexity. In the same spirit, HotStuff \cite{hotstuff} (a concurrent proposal) incurs  the same message complexity, sharing with Tendermint a linear proposer replacement. Note as well that the proposed methodology allowed us to identify bugs (see \cite{tendermintTRv1}) in the preliminary versions of the protocol  (\cite{BKM18,Tendermint}).

This paper and \cite{ADPT18} target two different consensus algorithms that are core of two different releases of Tendermint blockchain.
In \cite{ADPT18} the authors reverse-engineered and then formalized the Tendermint blockchain protocol implemented initially by the Tendermint Foundation \cite{github-tendermint}. \cite{ADPT18} allowed to identify several bugs in the initial version of Tendermint implementation (see \cite{tendermintTRv1}). Moreover, we proved that  the termination property cannot be guaranteed in general, and hence an additional assumption on the execution is needed to solve Consensus. After the publication of our findings, Tendermint foundation  proposed a new algorithm, \cite{BKM18}, that is currently  implemented as consensus-core for the new release of Tendermint. The new version of the protocol claimed to include new mechanisms that removed the need of additional assumptions in order to guarantee the termination.
The pseudo-code proposed in \cite{BKM18} and further implemented by Tendermint foundation still had some bugs at the time when we started to analyse it, which we reported \cite{githubissues}.

In order to help practitioners, and in particular Tendermint foundation, to detect easily their errors and compare with the existing state of the art, 
in this paper we decided to have a bottom up approach by identifying the minimal building blocks a PBFT-like protocol should include in order to solve consensus function on the considered system and communication model (going from synchronous to eventually synchronous) and the behavior of Byzantine nodes. 
We used Tendermint as case study and identified the mechanisms needed by the protocol in order to be correct. Our study  resulted in three variants of the protocol for which we analyzed the correctness and the complexity. In this paper, we included two of the three algorithms (we decided to left aside the trivial one where Byzantines have a symmetrical behavior and the communication is synchronous). Moreover, the complexity analysis proposed in our paper may help both practitioners and academics to compare Tendermint to the state of the art which was an open question so far.
\section{Model}\label{sec:model}

The system is composed of an infinite set $\Pi$ of sequential processes, namely $\Pi = \{p_1, \dots\}$; 
\textit{Sequential} means that a process executes one step at a time.
This does not prevent it from executing several threads with an appropriate multiplexing. As local processing time are negligible with respect to message transfer delays, they are considered as equal to zero.

\textbf{Arrival model.}
We assume  a \textit{finite arrival model} \cite{aguilera2004}, i.e.  the system has  infinitely many processes $\Pi$ but each run has only finitely many.   The size of the set $\Pi_{\rho} \subset \Pi$  of processes that participate  in each system run is not a priori-known. 
We also consider a finite subset $V \subseteq \Pi_\rho$  of validators. The set $V$ may change during any system run and its size $n$ is a-priori known. A process is promoted in $V$ based on a so-called merit parameter, which can model for instance its stake in proof-of-stake blockchains. Note that in the current Tendermint implementation, it is a separate module included in the Cosmos project \cite{cosmos} that is  in charge of implementing the selection of $V$.

\textbf{Failure model.} There is no bound on processes that can exhibit a Byzantine behaviour \cite{RA80} in the system, but up to $f$ validators can exhibit a Byzantine behaviour at each point of the execution. A Byzantine process is a process that behaves arbitrarily.
A process (or validator) that exhibits a Byzantine behaviour is called \textit{faulty}. Otherwise, it is \textit{non-faulty} or \textit{correct} or \textit{honest}.
To be able to solve the consensus problem, we assume that $f < n/3$ and more precisely we consider $n=3f+1$. 

\textbf{Communication model.} Processes communicate by exchanging messages through an eventually synchronous  network \cite{DLS88}. \textit{Eventually Synchronous} means that after a finite unknown time  $\tau>0$ there is a bound $\delta$ on the message transfer delay. When   $\tau=0$  the network is \emph{synchronous}.

In the following we assume the presence of a \emph{broadcast primitive}. A process $p_i$ by invoking the primitive $\textsf{broadcast} (\langle TAG,m \rangle)$ broadcasts a message, where $TAG$ is the type of the message, and $m$ its content. To simplify the presentation, it is assumed that a process can send messages to itself. The primitive $\textsf{broadcast} ()$ is a best effort broadcast, which means that when a correct process broadcasts a value, eventually all the correct processes deliver it.
A process $p_i$ receives a message by executing the primitive $\textsf{delivery}()$.
Messages are created with a digital signature, and we assume that digital signatures cannot be forged. When a process $p_i$ delivers a message, it knows the process $p_j$ that created the message.

Let us note that the assumed broadcast primitive in an open dynamic network can be implemented through $gossiping$, i.e. each process sends the message to current neighbors in the underlying dynamic network graph. In these settings the finite arrival model is a necessary condition for the system to show eventual synchrony. Intuitively, a finite arrival implies that message losses due to topology changes are bounded, so that the propagation delay of a message between two processes not directly connected can be bounded \cite{BBTR2007,MJ18}.


\textbf{Round-based execution model.} We assume that each correct process evolves in rounds.
A \emph{round} consists of three phases, in order: (i) a \emph{Send} phase, where the process broadcasts messages computed during the last round, or a default messages for the first round; (ii) a \emph{Delivery} phase where the process collects messages sent during the current and previous rounds; and (iii) a \emph{Compute} phase where the process uses the messages delivered to change its state.
At the end of a round a process exits from the current round and starts the next round.
Each round has a finite duration, we consider the Send and the Compute phase as being atomic, they are executed instantaneously, but not the Delivery phase.
In a synchronous network, we assume the duration of the Delivery phase, and so of the round is $\delta$. 
We assume that processes have no access to a global clock but have access to local clocks, these clocks might not be synchronized with each other but are allowed to have bounded clock skew.

\textbf{Problem definition.}
In this paper we analyze the correctness of Tendermint protocol with respect to the  consensus specification:
	\textbf{Termination}, every correct process eventually decides some value; \textbf{Integrity}, no correct process decides twice; \textbf{Agreement}, if there is a correct process that decides a value $v$, then eventually all the correct processes decide $v$; \textbf{Validity\cite{CKPS01,redbelly17}}, a decided value is valid, it satisfies the predefined predicate denoted $\textsf{valid}()$.

\section{Tendermint Algorithms}\label{sec:algo}

Tendermint BFT Consensus protocol  \cite{BKM18,Tendermint,github-tendermint} is a variant of PBFT consensus, at the core layer of the Tendermint blockchain.

The algorithm follows the rotating coordinator paradigm i.e., for each new block to be appended there is a proposer, chosen among the validators, that proposes the block. If the block is not decided then a new proposer is selected and so on, until a block is decided by all the correct validators and consensus terminates.
In the following we present variants of \cite{BKM18} in synchronous and eventual synchronous communication models.
\paragraph{Basic principles of the protocol.}
Each block in the blockchain is characterized by its height $h$, which is the distance in terms of blocks from the genesis block, which is at height $0$. For each new height, the two protocols (Algorithm \ref{alg:tendermintSync} for the synchronous case and Algorithm \ref{alg:tendermintCorrected} for the eventual synchronous case) share a common algorithmic structure, they proceed in \emph{epoch}s, and each epoch $e$ consists in three rounds: 
 the \emph{PRE-PROPOSE} round;  the \emph{PROPOSE} round; and the \emph{VOTE} round.
During the PRE-PROPOSE round, the proposer pre-proposes a value $v$ to all the other validators. During the PROPOSE round, if a validator accepts $v$ then it proposes such value. If a validator receives \emph{enough} proposals for the same value $v$ then it votes for $v$ during the VOTE round. Finally, if a validator receives \emph{enough} votes for $v$, it decides on $v$. In this case, \emph{enough} means at least $2f+1$ occurrences of the same value from $2f+1$ different validators and from each validator only the first value delivered for each round is considered, (cf. Algorithm \ref{alg:messages}).

If the proposer is correct then it pre-proposes the same value to all the $2f+1$ correct validators. All the $2f+1$ correct validators propose such value, it follows that all the $2f+1$ correct validators vote for such value and decide for it. If the proposer is Byzantine  it can pre-propose different values to different correct validators, creating a partition in the proposal value set collected by validators. Depending on what the remaining Byzantine validators do, some correct validators may decide on a value $v$ and some other may not\footnote{Since there are $3f+1$ validators, there cannot be two different values that collect $2f+1$ distinct votes in the same epoch.}, then a new epoch starts.  In order to not violate the agreement property, validators that have not decided yet in the previous epoch must only decide for $v$, for this reason validators, before vote for some value $v$, lock on that value, i.e., they will refuse to propose a further pre-proposed value different than $v$.

\paragraph{Information from one epoch to the next.} $lockedValue$ and $validValue$ variables\footnote{$validValue$ was not present in the previous version of Tendermint \cite{Tendermint}, that was suffering from the Live Lock bug \cite{livelock}.} carry the potentially decided value from one epoch to the next one.
The $lockedValue$ idea is the following. If one correct validator decides on $v$, it means that it collected $2f+1$ votes for $v$ during the VOTE phase, since there are at most $f$ Byzantine validators thus there are at least $f+1$ correct validators that voted for $v$ and those validators must not vote for any other different value than $v$. For this reason if a validator delivers $2f+1$ proposals for $v$ during the PROPOSE round it sets its $lockedValue$ to $v$. Since each new pre-proposed value $v'$ is proposed if $v'$ is equal to $lockedValue$ or $validValue$ (not true for at lest $f+1$ correct validators that set $lockedValue$ to $v$), then there can be at most $2f$ possible proposals for $v'$ that are not enough to lock and vote for $v'$, i.e., it is not possible to decide for any value different than $v$. On the other side, if no correct validator decided yet, Byzantine faulty validators may force different correct validators to lock on different values. Let us consider a scenario where the proposer is Byzantine and proposes $v$ to $f+1$ correct validators and then $f$ Byzantine validators make $x\leq f$ of them lock on $v$ and a similar scenario can happen with another value $v'$ so that we can have different correct validators, let us say $y\leq f$ locked on a different value. If any new pre-proposal is checked only against the $lockedValue$ then a correct validator locked on a value $v$ refuses (does not propose) all values different from $v$, it means that when some correct validator is locked, the proposer needs to propose some of the value on which the correct validators are locked on, but such value, in order to be accepted cannot be checked only against the $lockedValue$ because we may never have enough correct validators proposing such value. For this reason validators keep track of the $validValue$ and by construction of the algorithm all correct validators have the same $validValue$ at the end of the epoch (in the synchronous period). Such value is then used to set the value to pre-propose and it is further used along with $lockedValue$ to accept or not a pre-proposed value. 

\noindent
\textbf{Messages syntax.}
When the validator $p_i$ broadcasts a message $\langle TAG,h,e,m\rangle$, where $m$ contains a value $v$, we say that $p_i$ pre-proposes, proposes or votes $v$ if $TAG$=$\texttt{PRE-PROPOSE}$, $TAG$=$\texttt{PROPOSE}$, $TAG$=$\texttt{VOTE}$, respectively.
\\
\textbf{Variables and data structures.}
	$h$  is an integer representing the consensus instance the validator is currently executing.
	$e_i$  is an integer representing the epoch where the validator $p_i$ is, we note that for each height, a validator may have multiple epochs.
	$decision_i$ is the decision of validator $p_i$ for the consensus instance $h$.
	$proposal_i$ is the value the validator $p_i$ proposes.
	$vote_i$ is the value the validator $p_i$ votes.
	$lockedValue_i$ stores a value which is potentially decided by some other validator. If validator $p_i$ delivers more than $2f+1$ proposes for the same value $v$ during its PROPOSE round, it sets $lockedValue_i$ to $v$.
	$validValue_i$ stores a value which is potentially decided by some other validator. If the validator $p_i$ delivers at least $2f+1$ proposes for the same value $v$ (from different validators) whether during its PROPOSE round or its VOTE round, it sets $validValue_i$ to $v$. $validValid_i$ is the last value that a validator delivered at least $2f+1$ times, and can be different than $lockedValue_i$. The latter two variables are used as follows: if $p_i$ is the next proposer then $p_i$ pre-proposes $validValid_i$ if different from $nil$. Otherwise, if $p_i$ is a validator, it checks the new pre-proposal against $lockedValue_i$ and $validValid_i$ if those are different from $nil$.
\\
\textbf{Functions.}
We denote as $Value$ the set containing all blocks, as $MemPool$ the set containing all the transactions, and as $Messages$ the set containing all messages.\\
\noindent{\bf -} $\textsf{proposer}: Height \times Epoch \to V \subseteq \Pi_\rho $ is a deterministic function which gives the proposer out of the validators set for a given epoch at a given height in a round robin fashion.\\
\noindent{\bf -} $\textsf{valid}: Value \to Bool$ is an application dependent predicate that is satisfied if the given value is valid w.r.t. the blockchain. If there is a value $v$ such that $\textsf{valid} (v) = \textsf{true}$, we say that $v$ is valid. Note that we set $\textsf{valid}(\nil) = \textsf{false}$.\\
\noindent{\bf -} $\textsf{getValue}()$ return a valid value.\\ 
\noindent{\bf -} $\textsf{sendByProposer}: Height \times Epoch \times Value \to Bool$ is an predicate that gives $\textsf{true}$ if the given value has been pre-proposed by the proposer of the given height during the given epoch. \\
\noindent{\bf -} $2f+1: \mathcal{P}(\texttt{Messages}) \to Bool$: checks if there are at least $2f+1$ proposals (resp. votes) in the given set of messages.

Everything defined above is common to the two algorithms. In each section we specify the data structures relative to a specific version of the algorithm.\\

\begin{algorithm}[t] \def\baselinestretch{1} \scriptsize\raggedright
	 \begin{algorithmic}[1] \SHORTSPACE 
		
		\SPACE \UPON{$\li{\text{TYPE},h,\currEpoch, \text{message}}$ \From\ validator $p_j$} \label{line:tab:recvProposal}
			\IF{$\nexists c: (\li{\text{TYPE},h,\currEpoch, c},p_j)\in messagesSet$} \label{line:tab:acceptProposal1}
				\STATE $messagesSet_i \assign messagesSet_i \cup (\li{\text{TYPE},h,\currEpoch, \text{message}},p_j)$
			\ENDIF 
		\ENDUPON		
	\end{algorithmic} 
	\caption{Messages management for validator $p_i$}
	\label{alg:messages} 
\end{algorithm}

\subsection{Byzantine Synchronous System}\label{ssec:algoSynchronous}

\begin{algorithm}[tp] \def\baselinestretch{1} \scriptsize\raggedright
	 \begin{algorithmic}[1] \SHORTSPACE \INIT{} 
		\STATE $\currEpoch_i := 0$   \COMMENT{This current epoch number}
		\STATE $decision_i := nil$ \COMMENT{This variable stocks the decision of the validator $p_i$}
		\STATE $lockedValue_i := nil; validValue_i := nil$
		\STATE $proposal_i := getValue()$ \COMMENT{This variable stocks the value the validator will (pre-)propose}
		\STATE $v_i := \nil$ \COMMENT{Local variable stocking the pre-preposal if delivered}
		\STATE $vote_i := nil$ 
		\ENDINIT 
		
		\SPACE \ROUND{PRE-PROPOSE($\currEpoch_i$)}\label{Pprop-beginSync}
			\SENDP{}
				\IF{$decision_i \neq \nil$}\label{bcVote}
					\STATE $\forall v, p_j: (\li{\Precommit,h,e_i,v},p_j) \in messagesSet_i, \Broadcast \li{\Precommit,h,e_i,v}$
					\STATE \textsf{return}
				\ENDIF
				\IF{$\coord(h, \currEpoch_i) = p_i$}
					\STATE \Broadcast\ $\li{\Proposal,h, \currEpoch_i, proposal_i}$ to all validators
				\ENDIF
			\ENDSENDP
			\DEL{}
				\WHILE{$(\timerPrePropose \text{ not expired})$}
					\IF{$\exists v: \sendBy(h,\currEpoch_i,v)$}
						\STATE $v_i \assign v$ \COMMENT{$v$ is the value sent by the proposer}
					\ENDIF	
				\ENDWHILE
			\ENDDEL
			\COMP{}
				\IF{$!valid(v_i)$}
					\STATE $proposal_i \assign nil$ \COMMENT{Note that $valid(\nil)$ is set to $\textsf{false}$}
				 \ELSE \IF{$validValue_i= \nil \lor  v_i \in \{lockedValue_i, validValue_i\}$}
				 		\STATE $proposal_i \assign v_i$
				 	 \ELSE
				 		\STATE $proposal_i \assign nil$\label{Pprop-endSync}
				 	\ENDIF	
				\ENDIF
			\ENDCOMP
		\ENDROUND
		\end{algorithmic} 
	\caption{Simplified Algorithm part 1 for height $h$ executed at validator $p_i$}
	\label{alg:tendermintSync} 
\end{algorithm}

	In Algorithms \ref{alg:messages} - \ref{alg:tendermintSync2} we describe the algorithm to solve consensus in a synchronous system in presence of Byzantine failures. 	The algorithm proceeds in 3 rounds for any given epoch at height $h$:

	\begin{itemize}
		\item Round PRE-PROPOSE (lines \ref{Pprop-beginSync} - \ref{Pprop-endSync}, Algorithm \ref{alg:tendermintSync}):
			If the validator $p_i$ is the proposer of the epoch, it pre-proposes its proposal value,
			otherwise, it waits for the proposal from the proposer.
			The proposal value of the proposer is its $validValue_i$ if $validValue_i \neq \nil$.
			If a validator $p_j$ delivers the pre-proposal from the proposer of the epoch, $p_j$ checks the validity of the pre-proposal and if to accept it with respect to the values in $validValue_i$ and $lockedValue_i$. If the pre-proposal is accepted and valid, $p_j$ sets its proposal $proposal_j$ to the pre-proposal, otherwise it sets it to $\nil$.
\item Round PROPOSE (lines \ref{Prop-beginSync} - \ref{Prop-endSync}, Algorithm \ref{alg:tendermintSync2}):
			During the PROPOSE round, each validator broadcasts its  proposal, and collects the proposals sent by the other validators.
			After the Delivery phase, validator $p_i$ has a set of proposals, and checks if $v$, pre-proposed by the proposer, was proposed by at least $2f+1$ different validators, if it is the case, and the value is valid, then $p_i$ sets $vote_i, validValue_i$ and $lockedValue_i$ to $v$, otherwise it sets $vote_i$ to $nil$.
		\item Round VOTE (lines \ref{Vote-beginSync} - \ref{Vote-endSync}, Algorithm \ref{alg:tendermintSync2}):
			In the round VOTE, a correct validator $p_i$ votes $vote_i$ and broadcasts all the proposals it delivered during the current epoch.
			Then $p_i$ collects all the messages that were broadcast.
			First $p_i$ checks if it has delivered at least $2f+1$ of proposal for a value $v'$ pre-proposed by the proposer of the epoch, in that case, it sets $validValue_i$ to that value then it checks if a value $v'$ pre-proposed by the proposer of the current epoch is valid and has at least $2f+1$ votes, if it is the case, then $p_i$ decides $v'$ and goes to the next height; otherwise it increases the epoch number and updates the value of $proposal_i$ with respect to $validValue_i$.
	\end{itemize}
\begin{algorithm}[t] \def\baselinestretch{1} \scriptsize\raggedright
	\begin{algorithmic}[1] \SHORTSPACE 		
		\SPACE \ROUND{PROPOSE($\currEpoch_i$)}\label{Prop-beginSync}
		\SENDP{}
		\IF{$proposal_i \neq \nil$}
		\STATE \Broadcast \ $\li{\Prevote,h,\currEpoch_i,proposal_i}$ to all validators
		\ENDIF
		\ENDSENDP
		\DEL{}
		\STATE \textbf{while} $(\timerPropose \text{ not expires})$ \textbf{do}\{\} \COMMENT{Collect messages}
		\ENDDEL
		\COMP{}
		\IF{$\exists v: 2f+1 \li{\Prevote,h, \currEpoch_i,v} \land valid(v) \land \sendBy(h,\currEpoch_i,v)$}
			\STATE $lockedValue_i \assign v$
			\STATE $validValue_i \assign v$
			\STATE $vote_i \assign v$
			\ELSE
			\STATE $vote_i \assign \nil$\label{Prop-endSync}
		\ENDIF 
		\ENDCOMP
		\ENDROUND
		
		\SPACE \ROUND{VOTE($\currEpoch_i$)}\label{Vote-beginSync}
		\SENDP{}
		\STATE  $\forall v, p_j: (\li{\Prevote,h,e_i,v},p_j) \in messagesSet_i, \Broadcast \li{\Prevote,h,e_i,v}$
		\IF{$vote_i \neq \nil$}
			\STATE \Broadcast \ $\li{\Precommit,h,\currEpoch_i,vote_i}$ \label{Vote-bc}
		\ENDIF
		\ENDSENDP
		
		\DEL{}
		\STATE\textbf{while} $(\timerVote \text{ not expires})$ \textbf{do}\{\}\COMMENT{Collect messages}
		\ENDDEL
		
		\COMP{}
		\IF{$\exists v': 2f+1 \li{\Prevote,h, \currEpoch_i,v'} \land valid(v') \land \sendBy(h,\currEpoch_i,v')$}
		\STATE $validValue_i \assign v'$
		\ENDIF
		\IF{$\exists v_d, \currEpoch_d: 2f+1 \li{\Precommit,h,\currEpoch_d,v_d} \land valid(v_d) \land decision_i = \nil$}
			\STATE $decision_i \assign v_d$ 
		 \ELSE
		 \STATE $e_i \assign e_i+1$
		\STATE $v_i \assign nil$ 
		 \IF{$validValue_i \neq \nil$}
			\STATE $proposal_i \assign validValue_i$ 
			\ELSE 
			\STATE $proposal_i \assign getValue()$\label{Vote-endSync}
		 \ENDIF
		\ENDIF
		\ENDCOMP
		\ENDROUND
	\end{algorithmic} 
	\caption{Simplified Algorithm part 2 for height $h$ executed at validator $p_i$}
	\label{alg:tendermintSync2} 
\end{algorithm}

\begin{algorithm}[t] \def\baselinestretch{1} \scriptsize\raggedright
	 \begin{algorithmic}[1] \SHORTSPACE 
	 	\INIT{} 
			\STATE $\currEpoch_i := 0$   \COMMENT{Current epoch number}
			\STATE $decision_i := nil$ \COMMENT{This variable stocks the decision of the validator $p_i$}
			\STATE $lockedValue_i := nil$; $validValue_i := nil$
			\STATE $lockedEpoch_i := -1$; $validEpoch_i := -1$ 
			\STATE $proposal_i := getValue()$ \label{getVal} \COMMENT{This variable stocks the value the validator will (pre-)propose}
			\STATE $v_i := \nil$ \COMMENT{Local variable stocking the pre-preposal if delivered}
			\STATE $validEpoch_j := \nil$ \COMMENT{Local variable stocking the proposer's validEpoch}
			\STATE $vote_i := nil$    \COMMENT{This variable stock the value the validator will vote for}
			\STATE $\timeoutPropose := \Delta_{\text{Pre-propose}}$; $\timeoutPrevote := \Delta_{\text{Propose}}$; $\timeoutPrecommit := \Delta_{\text{Vote}}$
		\ENDINIT 
		
		\SPACE
		\ROUND{PRE-PROPOSE}\label{Pprop-beginESync}
			\SENDP{}
				\IF{$decision_i \neq \nil$}\label{bcVote}
					\STATE $\forall v, p_j: (\li{\Precommit,h,e_i,v},p_j) \in messagesSet_i, \Broadcast \li{\Precommit,h,e_i,v}$
					\STATE \textsf{return}
				\ENDIF	
				\IF{$\coord(h, \currEpoch_i) = p_i$}
					\STATE \Broadcast\ $\li{\Proposal,h, \currEpoch_i, proposal_i,validEpoch_i}$ \label{Pprop-bc}
				\ENDIF
			\ENDSENDP
			\DEL{}
				\STATE \textsf{set} $timerPrePropose$ {\sf to} $\timeoutPropose$
				\WHILE{$(\timerPrePropose \text{ not expired}) \land \lnot (\exists v_j,e_j: \sendBy(h,\currEpoch_i,v_j,e_j))$}
					\IF{$\exists v_j,e_j: \sendBy(h,\currEpoch_i,v_j,e_j)$}
						\STATE $v_i \assign v_j$ \COMMENT{$v_j$ is the value sent by the proposer}
						\STATE $validEpoch_j \assign e_j$ \COMMENT{$e_j$ is the $validEpoch$ sent by the proposer}
					\ENDIF	
				\ENDWHILE
				\IF{$\lnot (\exists v,epochProp: \sendBy(h,\currEpoch_i,v,epochProp))$}
					\STATE $\timeoutPropose \assign \timeoutPropose+1$
				\ENDIF
			\ENDDEL
			\COMP{}
				\IF{$2f+1$ $\li{\Prevote,h, validEpoch_j,v_i} \land validEpoch_j \ge lockedEpoch_i \wedge validEpoch_j < \currEpoch_i \wedge valid(v_i)$}\label{Pprop-prop}
					\STATE $proposal_i \assign v_i$
				\ELSE \IF{$!valid(v_i) \vee (lockedEpoch_i > validEpoch_j  \wedge lockedValue_i \neq v_i)$}
						\STATE $proposal_i \assign nil$ \label{Pprop-nil} \COMMENT{Note that $valid(\nil)$ is set to $\textsf{false}$}
					\ENDIF 
					 \IF{$valid(v_i) \wedge (lockedEpoch_i = -1  \vee lockedValue_i = v_i)$} 
						\STATE $proposal_i \assign v_i$\label{Pprop-endESync}
					
					\ENDIF
				\ENDIF
			\ENDCOMP
		\ENDROUND
	\end{algorithmic} 
	\caption{Tendermint Consensus part 1 for height $h$ executed by $p_i$}
	\label{alg:tendermintCorrected} 
\end{algorithm}
\begin{algorithm}[tp] \def\baselinestretch{1} \scriptsize\raggedright
	\begin{algorithmic}[1] \SHORTSPACE 
		\SPACE \ROUND{PROPOSE}\label{Prop-beginESync}
		\SENDP{}
		\IF{$proposal_i \neq \nil$}
			\STATE \Broadcast \ $\li{\Prevote,h,\currEpoch_i,proposal_i}$ \label{Prop-bc}
		\ENDIF
		\STATE \Broadcast\ $\li{\hb,\Prevote,h, \currEpoch_i}$
		\ENDSENDP
		\DEL{}
			\STATE \textsf{set} $\timerPropose$ {\sf to} $\timeoutPrevote$\label{Prop-del}
			\STATE\textbf{while} $(\timerPropose \text{ not expires}) \land \lnot (2f+1 \li{\hb,\Prevote,h, \currEpoch_i})$ \textbf{do}\{\} 
			\COMMENT{Note that the HeartBeat messages should be from different validators}
			\IF{$\lnot (2f+1 \li{\hb,\Prevote,h, \currEpoch_i})$}
				\STATE $\timeoutPrevote \assign \timeoutPrevote+1$
			\ENDIF
		\ENDDEL 
		\COMP{}
			\IF{$\exists v': 2f+1 \li{\Prevote,h, \currEpoch_i,v'} \land valid(v') \land \sendBy(h,\currEpoch_i,v') $}\label{Prop-maj}
				\STATE $lockedValue_i \assign v'$
				\STATE $lockedEpoch_i \assign \currEpoch_i$ \label{Prop-lock}
				\STATE $validValue_i \assign v'$
				\STATE $validEpoch_i \assign \currEpoch_i$
				\STATE $vote_i \assign v'$ \label{Prop-vote}
			\ELSE
				\STATE $vote_i \assign \nil$\label{Prop-endESync}
			\ENDIF 
		\ENDCOMP
		\ENDROUND
		
		\SPACE \ROUND{VOTE}\label{Vote-beginESync}
		\SENDP{}
		\STATE  $\forall v, p_j: (\li{\Prevote,h,e_i,v},p_j) \in messagesSet_i, \Broadcast \li{\Prevote,h,e_i,v}$
		\IF{$vote_i \neq \nil$}
			\STATE \Broadcast \ $\li{\Precommit,h,\currEpoch_i,vote_i}$ \label{Vote-bc}
		\ENDIF
		\STATE \Broadcast \ $\li{\hb,\Precommit,h, \currEpoch_i}$
		\ENDSENDP
		
		\DEL{}
		\label{Vote-del}
		\STATE \textsf{set} $\timerVote$ {\sf to} $\timeoutPrecommit$
		\STATE\textbf{while} $(\timerVote \text{ not expires}) \land \lnot (2f+1 \li{\hb,\Precommit,h, \currEpoch_i})$ \textbf{do}\{\} 
		\IF{$\lnot (2f+1 \li{\hb,\Precommit,h, \currEpoch_i})$}
		\STATE $\timeoutPrecommit \assign \timeoutPrecommit+1$
		\ENDIF
		\ENDDEL
		
		\COMP{}
		\IF{$\exists v'': 2f+1 \li{\Prevote,h, \currEpoch_i,v''}\land valid(v'') \land \sendBy(h,\currEpoch_i,v'') $}	
		\label{VOTE-validBegin}
		\STATE $validValue_i \assign v''$
		\STATE $validEpoch_i \assign \currEpoch_i$\label{VOTE-validEnd}
		\ENDIF
		\IF{$\exists v_d, \currEpoch_d: 2f+1 \li{\Precommit,h,\currEpoch_d,v_d}\land valid(v_d) \land decision_i = \nil$}\label{Vote-decideBegin}
		\STATE $decision_i \assign v_d$ \label{Vote-decide} \label{Vote-decideEnd}
		\ELSE\label{Vote-decideElse}
		\STATE $\currEpoch_i \assign \currEpoch_i+1$ \label{Vote-epoch}
		\STATE $v_i \assign \nil$
		\IF{$validValue_i \neq \nil$}
		\STATE $proposal_i \assign validValue_i$ 
		\ELSE 
		\STATE $proposal_i \assign getValue()$\label{Vote-endESync}
		\ENDIF
		\ENDIF
		\ENDCOMP
		\ENDROUND
	\end{algorithmic} 
	\caption{Tendermint Consensus part 2 for height $h$ executed by $p_i$}
	\label{alg:tendermintCorrected2} 
\end{algorithm}

\subsection{Byzantine Eventual Synchronous System}\label{ssec:algoEventualSynchronous}
This section presents the Algorithm \ref{alg:messages} and Algorithms \ref{alg:tendermintCorrected} - \ref{alg:tendermintCorrected2} that solve Consensus in an eventually synchronous model in presence of Byzantine faulty validators. This algorithm has been reported in an early version of \cite{BKM18} with the bugs fixed in \cite{githubissues}. To achieve the consensus in this setting two additional variables need to be used, (i) $lockedEpoch_i$ is an integer representing the last epoch where validator $p_i$ updated $lockedValue_i$, and (ii) $validEpoch_i$ is an integer which represents the last epoch where $p_i$ updates $validValue_i$.
These two new variables are used to not violate the agreement property during the asynchronous period. During such period different epochs may overlap at different validators, then it is needed to keep track of the relative epoch when a validator locks in order to not accept \vir{outdated} information generated during a previous epoch. Moreover, a round duration management mechanism needs to be introduced, i.e. increasing timeouts. In the previous algorithm, rounds were lasting $\delta$, the known message delay. In an eventually synchronous system such approach is not feasible, since during the asynchronous period messages may take unbounded delay before being delivered. It follows that, since there are at most $f$ Byzantine faulty validators, when a validator delivers messages from $n-f$ different validators it can terminate the delivery phase, but such phase may last an unbounded time. On the contrary, in the PRE-PROPOSE round only the proposer is sending a message, and generally messages may take a lot of time before being delivered, for such reasons timeouts need to be used in order to manage the rounds duration and adapted to  message delays, such that once the system enters in the synchronous period, rounds last enough for messages send during the round to be delivered before the end of it.

The algorithm proceeds in 3 rounds for any given epoch $e$ at height $h$. The description is mainly the same as in Section \ref{ssec:algoSynchronous}, thus in the following we underline just the differences:

\noindent {\bf -} Round PRE-PROPOSE (lines \ref{Pprop-beginESync} - \ref{Pprop-endESync}, Algorithm \ref{alg:tendermintCorrected}):
	The description of this round is mainly the same as before. We highlight the fact that a correct validator $p_i$ takes into account also $lockedEpoch_i$ in order to accept a pre-proposed value.\\
\noindent {\bf -} Round PROPOSE (lines \ref{Prop-beginESync} - \ref{Prop-endESync}, Algorithm \ref{alg:tendermintCorrected2}):
	When a correct validator $p_i$ updates $lockedValue_i$ (resp. $validValue_i$), it also update $lockedEpoch_i$ (resp. $validEpoch_i$) to the current epoch.\\
\noindent {\bf -} Round VOTE (lines \ref{Vote-beginESync} - \ref{Vote-endESync}, Algorithm \ref{alg:tendermintCorrected2}):
	If a correct validator $p_i$ delivered at least $f+1$ same type of messages from an epoch higher than the current one, $p_i$ moves directly to the PRE-PROPOSE round of that epoch and  when a correct validator $p_i$ updates $validValue_i$, it also update $validEpoch_i$ to the current epoch.

We recall that each validator has a time-out for each round. If during a round validator $p_i$ does not deliver at least $2f+1$ messages sent during that round (or the pre-proposal for the PRE-PROPOSE round), the corresponding time-out is increased. Those messages can be values or heartbeats, in the case in which a correct validator has not a value to propose or vote.
\subsection{Correctness Proof of Tendermint Algorithm in a Byzantine Eventual Synchronous Setting}		

In this section, we prove the correctness of $\Fig$ \ref{alg:tendermintCorrected} - \ref{alg:tendermintCorrected2} (Tendermint) in an eventual synchronous system. Due to the lack of space, the missing proofs can be found in the technical report \cite{dissectingTR}.

\begin{lemma}[Validity]\label{l:validity}
	In an eventual synchronous system, Tendermint verifies the following property:
	A decided value satisfies the predefined predicate denoted as $\textsf{valid}()$.
\end{lemma}


\begin{lemma}[Integrity]\label{l:integrity}
	In an eventual synchronous system, Tendermint verifies the following property:
	No correct validator decides twice.
\end{lemma}	

\begin{lemma}\label{l:decisionLock}
	Let $v$ be a value, $e$ an epoch, and the set $L^{v,e} = \{p_j: p_j \text{ correct} \land lockedValue_j = v \land lockedEpoch_j = e$ at the end of epoch $e$$\}$.
	In an eventual synchronous system, Tendermint verifies the following property: 
	If $|L^{v,e}| \ge f+1$ then no correct validator $p_i$ will have $lockedValue_i \neq v \land lockedEpoch_i \ge e$, at the end of each epoch $e' >e$, moreover a validator in $L^{v,e}$ only proposes $v$ or $\nil$ for each epoch $e' > e$.
\end{lemma}

\begin{lemma}[Agreement]\label{l:agreement}
	In an eventual synchronous system, Tendermint verifies the following property:
	If there is a correct validator that decides a value $v$, then eventually all the correct validators decide $v$.
\end{lemma}

\begin{lemma}[Termination]\label{l:termination}
	In an eventual synchronous system, Tendermint verifies the following property:
	Every correct validator eventually decides some value.
\end{lemma}

\begin{proofL}
	By construction, if a correct validator does not deliver more than $2f+1$ messages (or $1$ from the proposer in the PRE-PROPOSE round) from different validators during the corresponding round, it increases the duration of its round, so eventually during the synchronous period of the system all the correct validators will deliver the pre-proposal, proposals and votes from correct validators respectively during the PRE-PROPOSE, PROPOSE and the VOTE round.
	Let $e$ be the first epoch after that time.
	
	If a correct validator decides before $e$, by Lemma \ref{l:agreement} all correct validators decide which ends the proof.
	Otherwise at the beginning of epoch $e$, no correct validator decides yet.
	Let $p_i$ be the proposer of $e$. 
	We assume that $p_i$ is correct and pre-propose $v$; $v$ is valid since $getValue()$ always return a valid value (lines \ref{getVal}, $\Fig$ \ref{alg:tendermintCorrected} \& line \ref{Vote-endESync}, $\Fig$ \ref{alg:tendermintCorrected2}), and $validValue_i$ is always valid (lines \ref{Prop-maj} \& \ref{VOTE-validBegin}, $\Fig$ \ref{alg:tendermintCorrected2}). 
	We have 2 cases:
	\begin{itemize}
%
					
		\item{Case 1:} At the beginning of epoch $e$, $|\{p_j: p_j \text{ correct} \land (lockedEpoch_j \le validEpoch_i \lor lockedValue_j=v)\}| \ge 2f+1$.
					
					Let $p_j$ be a correct validator where the condition $lockedEpoch_j \le validEpoch_i \lor$ $lockedValue_j=v$ holds.
					After the delivery of the pre-proposal $v$ from $i$, $p_j$ will update $proposal_j$ to $v$ (lines \ref{Pprop-prop} - \ref{Pprop-endESync}, $\Fig$ \ref{alg:tendermintCorrected}). During the PROPOSE round, $p_j$ proposes $v$ (line \ref{Prop-bc}, $\Fig$ \ref{alg:tendermintCorrected2}), and since there are at least $2f+1$ similar correct validators they will all propose $v$, and
					all correct validators will deliver at least $2f+1$ proposals for $v$ (line \ref{Prop-del}, $\Fig$ \ref{alg:tendermintCorrected2}).
					
					Correct validators will set their $vote$ to $v$ (lines \ref{Prop-maj} - \ref{Prop-bc}, $\Fig$ \ref{alg:tendermintCorrected2}), will vote $v$, and will deliver these votes, so at least $2f+1$ of votes (lines \ref{Vote-bc} \& \ref{Vote-del}, $\Fig$ \ref{alg:tendermintCorrected2}). Since we assume that no correct validators decided yet, and since they deliver at least $2f+1$ votes for $v$, they will decide $v$ (lines \ref{Vote-decideBegin} - \ref{Vote-decide}, $\Fig$ \ref{alg:tendermintCorrected2}).
		\item{Case 2:} At the beginning of epoch $e$, $|\{p_j: p_j \text{ correct} \land (lockedEpoch_j \le validEpoch_i \lor lockedValue_j=v)\}| < 2f+1$.
					
					Let $p_j$ be a correct validator where the condition $lockedEpoch_j > validEpoch_i \land$ $lockedValue_j\neq v$ holds. When $p_i$ will make the pre-proposal, $p_j$ will set $proposal_j$ to $\nil$ (line \ref{Pprop-nil}, $\Fig$ \ref{alg:tendermintCorrected}) and will propose $\nil$ (line \ref{Prop-bc}, $\Fig$ \ref{alg:tendermintCorrected2}).
					
					By counting only the propose value of the correct validators, no value will have at least $2f+1$ proposals for $v$.
					There are two cases:
					\begin{itemize}
						\item No correct validator delivers at least $2f+1$ proposals for $v$ during the PROPOSE round,
							so they will all set their $vote$ to $\nil$, vote $\nil$ and go to the next epoch without changing their state (lines \ref{Prop-endESync} \& \ref{Vote-bc} - \ref{Vote-del} \& \ref{Vote-decideElse} - \ref{Vote-endESync}, $\Fig$ \ref{alg:tendermintCorrected2}).
							
						\item If there are some correct validators that delivers at least $2f+1$ proposals for $v$ during the PROPOSE round, which means that some Byzantine validators send proposals for $v$ to those validators.
							
							As in the previous case, they will vote for $v$, and since there are $2f+1$ of them, all correct validators will decide $v$. Otherwise, there are less than $2f+1$ correct validators that deliver at least $2f+1$ proposals for $v$. Only them will vote for $v$ (line \ref{Vote-bc}, $\Fig$ \ref{alg:tendermintCorrected2}). Without Byzantine validators, there will be less than $2f+1$ vote for $v$, no correct validator will decide (lines \ref{Vote-decideBegin} - \ref{Vote-decideEnd}, $\Fig$ \ref{alg:tendermintCorrected2}) and they will go to the next epoch, if Byzantine validators send votes for $v$ to a correct validator such as it delivers at least $2f+1$ votes for $v$ during VOTE round, then it will decide (lines \ref{Vote-decideBegin} - \ref{Vote-decide}, $\Fig$ \ref{alg:tendermintCorrected2}), and by Lemma \ref{l:agreement} all correct validators will eventually decide.
							
							Let $p_k$ be one of the correct validators that delivers at least $2f+1$ proposals for $v$ during PROPOSE round, it means that $lockedValue_{k}=v$ and $lockedEpoch_{k}=e$. It follows that  
							at the end of epoch $e$, all correct validators will have $validValue = v$ and $validEpoch = e$.
					\end{itemize}
					If there is no decision, either no correct validator changes its state, otherwise all correct validators change their state and have the same $validValue$ and $validEpoch$, eventually a proposer of an epoch will satisfy the case 1, and that ends the proof.
	\end{itemize}
	If $p_i$, the proposer of epoch $e$, is Byzantine and more than $2f+1$ correct validators delivered the same message during PRE-PROPOSE round, and the pre-proposal is valid, the situation is like $p_i$ was correct.
	Otherwise, there are not enough correct validators that delivered the pre-proposal, or if the pre-proposal is not valid, then there will be less than $2f+1$ correct validators that will propose that value, which is similar to the case 2.
	
	Since the proposer is selected in a round robin fashion, a correct validator will eventually be the proposer, and correct validators will decide.
	\renewcommand{\toto}{l:termination}
\end{proofL}

\begin{theorem}\label{t:consensus}
	In an eventual synchronous system, 
	Tendermint implements the consensus specification.
\end{theorem}

\subsection{Complexity of Tendermint Algorithm in a Byzantine Eventual Synchronous Setting}

Let us consider the following scenario after the asynchronous period (i.e., after $\tau$), in which in the first $f$ epochs, $e_{i+1}, \dots, e_{i+f}$, there are $f$ Byzantine proposers that make lock only one correct validator at each epoch on $f$ different values with different $lockedEpoch$, $e_{i+1}, \dots, e_{i+f}$. Let $p_j$ be the last correct validator that locked, and let $v$ such value ($lockedValue_j=v$) with $lockedEpoch_j=e_{i+f}$. Then all the other correct validators have $validValue$ set to $v$ and $validEpoch$ set to $e_{i+f}$. This happens thanks to the fact that when a correct validator locks on a value then at the end of the epoch every correct validator sets its $validValue$ to that value. The algorithm terminates when a pre-proposal is proposed and voted by more than $2f$ correct validators, i.e, when the pre-proposed value has $validEpoch$ greater equal than the validator $lockedEpoch$. Thus, during the period of synchrony, the first correct proposer that proposes leads the algorithm to terminate in $f+1$ rounds.
Let us consider the case in which there $f$ correct validators locked on $f$ different values with different $lockedEpoch$ before $\tau$. Let us assume that $p_j$ is the last correct validator that locked on a value $v$, thus it has the highest $lockedEpoch$ but not all the correct validators have their $validValue$ set to $v$ (due to the asynchronous communication). 
Let us now consider that after $\tau$ the first $f$ proposers are Byzantines and stay silent. The following proposers are correct but their pre-propose value might not be accepted by enough correct validators as long as $p_j$, with the highest $validEpoch$ and $lockedEpoch$ proposes. Which eventually happens due to the round robin selection function. Thus, the protocol terminates in a number of epochs proportional to the number of validators $O(n)$, while the lower bound to solve BFT Consensus in the worst case scenario is $f+1$ \cite{fischer1982lower}.
As for message complexity, since at each epoch, all validators broadcast messages, it follows that during one epoch the protocol uses $O(n^2)$ messages, thus in the worst case scenario the message complexity is $O(n^3)$.
 
 In the following we address the bit complexity of Tendermint.
In Tendermint, each message is composed as follow: 
\begin{itemize}
	\item \texttt{PRE-PROPOSE}: The marker that the message is from the round \texttt{PRE-PROPOSE}; two integers one for the current height, and the second for the current epoch; the proposed value; and an integer representing the epoch on which the proposer last updated   its $validValue$.
	\item \texttt{PROPOSE}: The marker that the message is from the round \texttt{PROPOSE}; two integers representing the current height and the current epoch; and a value which is the proposed block.
	\item \texttt{VOTE}: The marker that the message is from the round \texttt{VOTE}; two integers representing the current height and the current epoch; and a value which is the voted block.
	\item $\hb$: The marker that the HeartBeat is from the round \texttt{VOTE} or \texttt{PROPOSE}; two integers representing the current height and the current epoch.
\end{itemize}
A correct validator keeps in memory, for each epoch for a given height, one message for each type (\texttt{PROPOSE}, \texttt{VOTE}) and at most $2$ messages of type $\hb$ from each validator, and only one \texttt{PRE-PROPOSE}.
A correct validator may have at most $1$ message from \texttt{PRE-PROPOSE}, $n$ messages from \texttt{PROPOSE}, $n$ messages from \texttt{VOTE}, and $2n$ messages of type $\hb$.
Hence, for each epoch at any given height, a validator stores at most 4n+1 messages of size $O(\log{n})$.  In the worst case, for the whole execution, a validator may store $O(n^2)$ messages. Therefore, the bit complexity in the worst case is   $O(n^2\log{n})$.
 
 Note that  \cite{KM13} proposes a bit complexity of $O(n^3\log{n})$ for an optimal round complexity using a variant of the tree structure of the Exponential Information Gathering protocol introduced in \cite{GM93}.  
 Clearly, there is a tradeoff between the bit complexity and the round complexity of the Byzantine agreement.  
\section{Conclusion}\label{sec:conclusion}

The contribution of this work is twofold. First, it analyzes Tendermint consensus protocol and provides detailed proof of its correctness and complexity. Second, it dissects such protocol in order to  link   the algorithmic techniques to the considered system model. We believe that this methodology can contribute in making Byzantine-tolerant consensus algorithms more understandable for developers and practitioners.
\subsection*{Acknowledgment}
The authors would like to thank the reviewers of NETYS 2019 for their insightful comments.
The authors also thank Zaynah Dargaye for numerous discussions, and in particular for the consistency of this work.

\bibliographystyle{splncs04}
\bibliography{biblio,mariapotop}
\end{document}